\documentclass[review]{elsarticle}

\usepackage{lineno}
\usepackage[colorlinks,citecolor=blue,linkcolor=red,urlcolor=blue]{hyperref}
\usepackage{color}
\usepackage{fourier}
\usepackage{verbatim}
\usepackage{graphicx}
\usepackage{subfigure}
\usepackage{booktabs}
\usepackage{amsfonts}
\usepackage{multirow}
\usepackage{diagbox}
\usepackage{amsmath,amssymb}

\modulolinenumbers[5]

\journal{Int. J Heat Mass Transf.}

\begin{document}

\begin{frontmatter}

\title{Many-body effective thermal conductivity in phase-change nanoparticle chains due to near-field radiative heat transfer}

\author[mymainaddress,mysecondaddress]{Minggang Luo}

\author[mymainaddress,mythirdaddress]{Junming Zhao\corref{mycorrespondingauthor}}
\cortext[mycorrespondingauthor]{Corresponding author}
\ead{jmzhao@hit.edu.cn}

\author[myforthaddress]{Linhua Liu}

\author[mysecondaddress]{Brahim Guizal}

\author[mysecondaddress,myfifthaddress]{Mauro Antezza\corref{mycorrespondingauthor}}
\ead{mauro.antezza@umontpellier.fr}

\address[mymainaddress]{School of Energy Science and Engineering, Harbin Institute of Technology, 92 West Street, Harbin 150001, China}
\address[mysecondaddress]{Laboratoire Charles Coulomb (L2C) UMR 5221 CNRS-Universit\'e de Montpellier, F- 34095 Montpellier, France}
\address[mythirdaddress]{Key Laboratory of Aerospace Thermophysics, Ministry of Industry and Information Technology, Harbin 150001, China}
\address[myforthaddress]{School of Energy and Power Engineering, Shandong University, Qingdao 266237, China}
\address[myfifthaddress]{Institut Universitaire de France, 1 rue Descartes, F-75231 Paris Cedex 05, France}

\begin{abstract}
In dense systems composed of numerous nanoparticles, direct simulations of near-field radiative heat transfer (NFRHT) require considerable computational resources. NFRHT for the simple one-dimensional nanoparticle chains embedded in a non-absorbing host medium is investigated from the point of view of the continuum by means of an approach combining the many-body radiative heat transfer theory and the Fourier law. Effects of the phase change of the insulator-metal transition material (VO$_2$), the complex many-body interaction (MBI) and the host medium relative permittivity on the characteristic effective thermal conductivity (ETC) are analyzed. The ETC for VO$_2$ nanoparticle chains below the transition temperature can be as high as 50 times of that above the transition temperature due to the phase change effect. The strong coupling in the insulator-phase VO$_2$ nanoparticle chain accounts for its high ETC as compared to the low ETC for the chain at the metallic phase, where there is a mismatch between the characteristic thermal frequency and resonance frequency. The strong MBI is in favor of the ETC. For SiC nanoparticle chains, the MBI even can double the ETC as compared to those without considering the MBI effect. For the dense chains, a strong MBI enhances the ETC due to the strong inter-particles couplings. When the chains go more and more dilute, the MBI can be neglected safely due to negligible couplings. The host medium relative permittivity significantly affects the inter-particles couplings, which accounts for the permittivity-dependent ETC for the VO$_2$ nanoparticle chains. 
\end{abstract}

\begin{keyword}
effective thermal conductivity\sep near-field radiative heat transfer\sep many-body interaction \sep insulator-metal phase-change material\sep nanoparticles
\end{keyword}

\end{frontmatter}



\section{Introduction}
Near-field radiative heat transfer (NFRHT) is currently attracting a lot of interests for its fundamental and applicative facets \cite{Biehs2020review,Ben2019,DeSutter2019,Shen2020jap,Gelais2014}. In dense particulate systems, the separation distance between two nanoparticles is often comparable to or less than the characteristic thermal wavelength \cite{Luo2019}. Due to the near-field effect (e.g., evanescent wave tunneling), the heat flux will exceed the Planck\textquotesingle s blackbody limit by several orders of magnitude, which has been predicted thanks to the fluctuational electrodynamics theory \cite{Rytov1989,Polder1971,Volokitin2004,Narayanaswamy2008,Zheng2011IJHMT,Liu2014IJHMT} and proved by recent experimental observations \cite{Lim2020PRA,Sabbaghi2020jap,Ghashami2018,Song2015,Shen2009,Rousseau2009}.

Many important progresses have been reported on direct simulations of NFRHT for systems composed of nanoparticles. On the one hand, the inter-ensemble NFRHT between two nanoparticle ensembles (e.g., three-dimensional (3D) clusters of hundreds of nanoparticles \cite{Luo2019,Dong2017JQ,Chen2018JQ} and two-dimensional (2D) nanoparticle ensembles \cite{Luo2020,Phan2013,Luo2020apl}, as well as the simple ensembles composed of only a few nanoparticles \cite{Ben2011,DongPrb2017}) has been analyzed. On the other hand, the intra-ensemble NFRHT inside the nanoparticle ensemble itself has also been reported. The fractional diffusion theory was applied to describe NFRHT along 3D plasmonic nanostructure networks, as well as one-dimensional (1D) ones, and the heat superdiffusion behavior was found \cite{Ben2013}. For 2D fractal structures, the effects of the structure morphology on the collective properties were analyzed and the heat flux has no large-range character, in contrast to non-fractal structures \cite{Nikbakht2017}. Thermal radiation behavior along a 1D nanoparticle chain has been shown to be significantly affected by another nanoparticle chain in proximity due to strong couplings \cite{Luo2019JQ}. Radiative heat flux along a linear chain considering an external magnetic field was also analyzed \cite{Latella2017prl}. In addtion, thermal transport behaviors along the atomic chains due to quantum effects have been reported recently \cite{Doyeux2017,Doyeux2017prl,Leggio2015}. In general, the direct simulation of NFRHT for dense particulate systems composed of hundreds of thousands of nanoparticles will introduce considerable unknowns, which will be very time-consuming and will require considerable computational resources.

From the point of view of continuum, the method applying the effective thermal conductivity (ETC) to characterize the NFRHT in dense particulate systems based on the diffusion assumption is really time-saving, as compared to direct simulation methods (e.g., many-body radiative heat transfer theory \cite{Ben2011,DongPrb2017},  scattering matrix method \cite{Messina2011,Messina2014}, trace formulas method \cite{Kruger2012,Muller2017}, thermal discrete dipole approximation method (T-DDA) \cite{Edalatpour2014JQ,Edalatpour2015PRE,Edalatpour2016JQ}, fluctuating surface currents approach (FSC)\cite{Rodriguez2012FSC}, boundary element method (BEM) \cite{Reid2011BEM}, finite difference time domain method (FDTD) \cite{Rodriguez2011FDTD,Datas2013FDTD,Didari2014JQ} and the quasi-analytic solution \cite{Czapla2019}, to name a few). The kinetic theory (KT) framework was applied to obtain the ETC for 1D nanoparticle chains \cite{Ben2008,Tervo2016}. Recently, the limitations of KT framework to describe NFRHT was analyzed systematically: 1) the KT framework is not suitable for materials with resonant modes outside the Planck\textquotesingle s window (e.g., metal Ag) and 2) the KT framework cannot be applied directly to 2D and 3D systems due to the lack of dispersion relations for these systems \cite{Kathmann2018,Tervo2020}. Most recently, a new method based on the many-body radiative heat transfer theory and the Fourier law (MF method) was proposed by Tervo \textit{et al}. \cite{Tervo2019} to obtain the ETC for arbitrary nanoparticle collections allowing the comparison between different materials and different heat transfer modes.  

Due to the limitation of the KT framework describing ETC for nanoparticle ensembles NFRHT, the investigation on ETC for materials supporting resonances outside the Planck\textquotesingle s window is still missing. Furthermore, VO$_2$ attracts lots of interests because of its special insulator-metal transition behavior around its phase transition temperature. Besides, there is still lack of the investigation on the ETC for phase-change VO$_2$ nanoparticle ensembles, especially for its metallic phase. Based on the promising phase-changing characteristics for VO$_2$, many potential applications were proposed recently: 1) near-field applications (e.g., the radiative thermal rectifier \cite{Zhang2020IJHMT,Zheng2017}, thermal transistor \cite{Biehs2014}, conductive thermal diode \cite{Ordonez-Miranda2018diode}, dynamic radiative cooling \cite{Taylor2017} and the scalable radiative thermal logic gates \cite{Kathmann2020}); 2) far-field applications (e.g., radiative thermal memristor \cite{Ordonez-Miranda2019} and radiative thermal rectifier \cite{Ordonez-Miranda2018}). Hence, it\textquotesingle s worth analyzing the effects of the phase change on the ETC for VO$_2$ nanoparticle ensembles, in addition to the phase-change effect on the thermal conductance reported very recently \cite{Luo2020}.

Nanoparticles in a dense particulate system often lie in the near field of each other, which results in the many-body interaction (MBI) making the NFRHT mechanism more complex \cite{Ben2011,Tervo2019,Tervo2017}. Though the complex MBI effects on the radiative heat flux and thermal conductance for various nanoparticle systems (three-nanoparticle system \cite{Ben2011,Wang2016,Song2019}, 1D nanoparticle chains \cite{Luo2019JQ}, 2D nanoparticle ensembles \cite{Luo2020,Luo2020apl,Nikbakht2017}, 3D nanoparticle ensembles \cite{Luo2019,Dong2017JQ,Chen2018JQ}) have been analyzed, the MBI effect on the ETC is still missing.

We extract simple 1D nanoparticle chains from realistic 3D nanoparticle ensembles embedded in a non-absorbing host medium and focus on the thermal property (i.e., effective thermal conductivity) describing and characterizing the NFRHT from the point of view of continuum. The relative permittivity of the host medium significantly affects the inter-particle couplings \cite{Tervo2020,Tervo2017}. Effects of the host medium relative permittivity on the ETC of closely spaced 1D metallic nanoparticle chains have already been analyzed \cite{Ben2008}. However, effects of the host medium relative permittivity on the ETC for the specific case of phase-change VO$_2$ nanoparticles have not yet been investigated.

We address the aforementioned missing points in this paper, where the ETC for the 1D nanoparticle chains of interest is obtained by means of the MF method. In Sec.~\ref{Models}, we give a brief description of the theoretical models for the MF method, as well as the formulas concerning the ETC for 1D nanoparticle chains. In Sec.~\ref{Results}, effects of the phase change of the insulator-metal transition material (i.e., VO$_2$), complex many-body interaction and host medium relative permittivity ($\epsilon_{\rm m}$) on the effective thermal conductivity due to NFRHT are analyzed. The optical properties for the materials used in this work are also given in this section.

\section{Models}
\label{Models}

\begin{figure} [htbp]
\centerline {\includegraphics[width=0.7\textwidth]{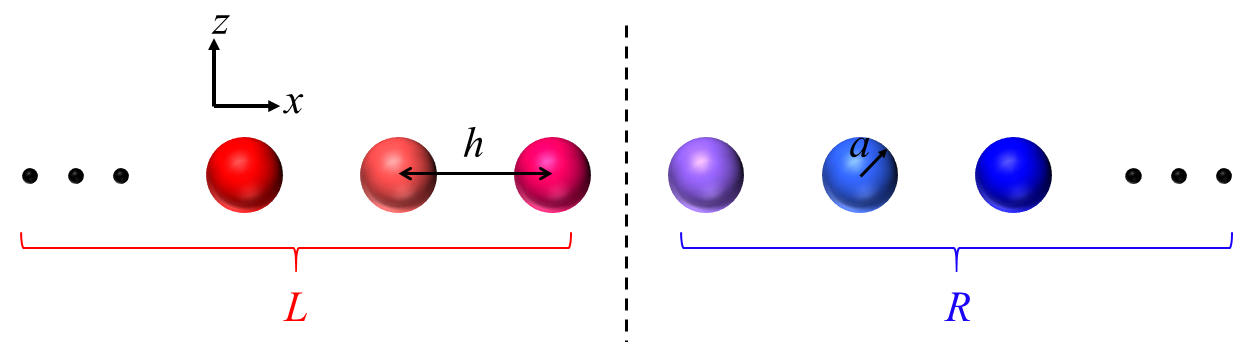}}
\caption{Schematic of the ordered nanoparticle chain embedded in a non-absorbing host medium with permittivity $\epsilon_{\rm m}$. The radiative heat flux in the chain is the sum of the net heat exchange between nanoparticles from part $L$ and nanoparticles from part $R$. Parts $L$ and $R$ are separated by the imaginary surface (dash line). The lattice spacing is $h$. Nanoparticle radius is $a$. A small linear temperature gradient d$T$/d$x$ along the chain is assumed. All particles are assumed near thermal equilibrium.}
\label{Structure_diagram}
\end{figure}

In this section, we describe in brief the physical system (the schematic is shown in Fig.~\ref{Structure_diagram}) and the theoretical aspects of the MF method for the ETC due to NFRHT in 1D nanoparticle chains. The nanoparticle chain is divided into two parts $L$ and $R$ by an imaginary plane.

The ETC due to near-field radiative heat transfer is defined as follows:
\begin{equation}
k_{eff}=\frac{Q}{A\cdot|\textrm{d}T/\textrm{d}x|},
\label{ETC}
\end{equation}
where $A$ is the cross section, $Q$ is the net radiative heat flux in the chain near thermal equilibrium with a small linear temperature gradient d$T$/d$x$, which is given as follows \cite{Tervo2019}.
\begin{equation}
Q=\sum_{\textrm{i}\in L}\sum_{\textrm{j}\in R}G_{\textrm{ij}}(T)d_{\textrm{ij}}\left|\frac{\textrm{d}T}{\textrm{d}x}\right|,
\label{power}
\end{equation}
where $d_\textrm{ij}$ is the separation distance center to center between the nanoparticle i from the part $L$ of the chain and nanoparticle j from the part $R$ of the chain, $G_\textrm{ij}(T)$ is the radiative thermal conductance between nanoparticles i and j, which yields \cite{Luo2020apl,Ben2013}
\begin{equation}
G_{\textrm{ij}}(T)=3\int_0^{+\infty}\frac{\textrm{d}\omega}{2\pi}\frac{\partial \Theta (\omega,T)}{\partial T}\mathcal{T}_{\rm i,j}(\omega),
\label{conductance}
\end{equation}
where $\Theta (\omega,T)$ is the mean energy of a harmonic Planck\textquotesingle s oscillator, $\omega$ is the angular frequency, the parameter $\mathcal{T}_{\rm i,j}(\omega)=\frac{4}{3}\frac{k^4}{\epsilon_{\rm m}}{\rm Im}(\chi_E^{\rm i}){\rm Im} (\chi_E^{\rm j})\textrm{Tr}(G_{\rm ij}^{EE}G_{\rm ij}^{EE\dagger})$, $\chi_{E}^{}=\alpha_E^{}-\frac{ik^3}{6\pi}|\alpha_E^{}|^2$ , $\alpha_E^{}$ is the nanoparticle polarizability, $k=\sqrt{\epsilon_{\rm m}}~\omega/c$ is the wave vector in the host medium, $\epsilon_{\rm m}$ is the host medium relative permittivity, $c$ is the speed of light in vacuum, the Green \textquotesingle s function $G_{\rm ij}^{EE}$ in the many-particle system naturally includes the many-body interaction and is the element of the following left matrix.
\begin{equation}
\begin{pmatrix}
0 & G_{12}^{EE} & \cdots & G_{1N}^{EE}\\
G_{21}^{EE} & 0 & \ddots & \vdots\\
\vdots & \vdots & \ddots & G_{(N-1)N}^{EE}\\
G_{N1}^{EE} & G_{N2}^{EE} & \cdots & 0
\end{pmatrix}=\begin{pmatrix}
0 & G_{0,12}^{EE} & \cdots & G_{0,1N}^{EE}\\
G_{0,21}^{EE} & 0 & \ddots & \vdots\\
\vdots & \vdots & \ddots & G_{0,(N-1)N}^{EE}\\
G_{0,N1}^{EE} & G_{0,N2}^{EE} & \cdots & 0
\end{pmatrix}\mathbb{A}^{-1},
\label{Green_function_MBI}
\end{equation}
where $G_{\rm 0,ij}^{EE}=\frac{e^{ikr}}{4\pi r}\left[\left(1+\frac{ikr-1}{k^{2}r^{2}}\right)\mathbb{I}_{3}+\frac{3-3ikr-k^{2}r^{2}}{k^{2}r^{2}}\hat{\textbf{r}}\otimes\hat{\textbf{r}}\right]$ is the free space Green\textquotesingle s function connecting two nanoparticles at $\textbf{r}_{\rm i}^{}$ and $\textbf{r}_{\rm j}^{}$, $r$ is the magnitude of the separation vector $\textbf{r}=\textbf{r}_{\rm i}^{}-\textbf{r}_{\rm j}^{}$, $\hat{\textbf{r}}$ is the unit vector $\textbf{r}/r$, $\mathbb{I}_{3}$ is the $3\times3$ identity matrix  and the matrix $\mathbb{A}$ including many-body interactions is defined as
\begin{equation}
\mathbb{A}=\mathbb{I}_{3N}^{}-k^2\begin{pmatrix}
0 & \alpha_{E}^{1}G_{0,12}^{EE} & \cdots & \alpha_{E}^{1}G_{0,1N}^{EE}\\
\alpha_{E}^{2}G_{0,21}^{EE} & 0 & \ddots & \vdots\\
\vdots & \vdots & \ddots & \alpha_{E}^{N-1}G_{0,(N-1)N}^{EE}\\
\alpha_{E}^{N}G_{0,N1}^{EE} & \cdots & \alpha_{E}^{N}G_{0,N(N-1)}^{EE} & 0
\end{pmatrix},
\label{matrix_interaction}
\end{equation} 
where $\mathbb{I}_{3N}$ is the $3N\times 3N$ identity matrix. Hence, the effective thermal conductivity will be rearranged as
\begin{equation}
k_{eff}=\frac{1}{A}\sum_{\textrm{i}\in L}\sum_{\textrm{j}\in R}G_{\rm ij}^{}(T)d_{\rm ij}^{}.
\label{ETC_rearranged}
\end{equation}
The effective thermal conductivity $k_{eff}$ can also be expressed as the frequency integral of the spectral effective thermal conductivity $k_{\omega}$: $k_{eff}=\int_{0}^{+\infty} k_{\omega}~\text{d}\omega$. For materials (e.g., metal Ag) where the magnetic-magnetic polarized eddy-current Joule dissipation dominates the radiative heat transfer, rather than the electric-electric polarized displacement current dissipation, the magnetic dipole contribution to the radiative heat transfer can be taken into consideration in the parameter $\mathcal{T}_{\rm i,j}(\omega)$ by the coupled electric and magnetic dipole approach \cite{Luo2019,Luo2020}.

\section{Results and discussion}
\label{Results}
In this section, the optical properties of phase-change VO$_2$ and polar SiC nanoparticles are introduced. Effects of the phase change, complex many-body interaction and host medium relative permittivity on the ETC of the 1D nanoparticle chains due to the NFRHT are analyzed by means of the MF method. We consider particles with radius $a=25$ nm forming a chain with 250 elements in part $L$ and as many in part $R$. This is large enough to reach convergent results for all the calculations of interest considered here \cite{Tervo2020}.
\subsection{Dielectric function and polarizability of nanoparticles}
VO$_2$ is a kind of phase-change materials, which undergoes an insulator-metal transition around 341 K (phase transition temperature). Below 341 K, VO$_2$ is an uniaxial anisotropic insulator, of which the dielectric function can be described by the following tensor \cite{Zhang2020IJHMT,Zheng2017,VO21966}:
\begin{equation}
\begin{pmatrix}
\epsilon_\rVert^{} & 0 & 0\\
0 & \epsilon_\bot^{} & 0\\
0 & 0 & \epsilon_\bot^{}
\end{pmatrix},
\label{dielectric_tensor}
\end{equation}
where $\epsilon_\bot^{}$ and $\epsilon_\rVert^{}$ are the ordinary and extraordinary dielectric function component relative to the optical axis, respectively. Both $\epsilon_\bot^{}$ and $\epsilon_\rVert^{}$ can be described by the Lorentz model as follows:
\begin{equation}
\epsilon(\omega)=\epsilon_{\infty}^{}+\sum_{n=1}^{N_L}\frac{S_n\omega_n^2}{\omega_n^2-i\gamma_n^{}\omega-\omega^2},
\label{Lorentz_model_VO}
\end{equation}
where $S_n$, $\omega_n^{}$ and $\gamma_n^{}$ are the phonon strength, phonon frequency and damping coefficient of the $n^{th}$ phonon mode. $N_L$ is the number of phonon modes ($N_L=8$ for $\epsilon_\bot^{}$ and $N_L=9$ for $\epsilon_\rVert^{}$). All the necessary parameters for both $\epsilon_\bot^{}$ and $\epsilon_\rVert^{}$ can be found in Ref.\cite{VO21966}. Above 341K, the dielectric function of the metallic-phase VO$_2$ is described by the Drude model as follows \cite{Zhang2020IJHMT,Zheng2017,VO21966}:
\begin{equation}
\epsilon(\omega)=-\epsilon_{\infty}^{}\frac{\omega_p^2}{\omega^2+i\omega\gamma},
\label{Drude_model_VO}
\end{equation}
where $\epsilon_{\infty} = 9$, $\omega_p$ = $1.51\times10^{15}$ rad$\cdot$s$^{-1}$ and $\gamma$ =$1.88\times10^{15}$ rad$\cdot$s$^{-1}$. In addition to the phase-change VO$_2$, the polar SiC is also used. The dielectric functions of SiC is described by the Drude-Lorentz model $\epsilon(\omega) =\epsilon_{\infty}^{}(\omega^2-\omega_l^2+i\gamma\omega)/(\omega^2-\omega_t^2+i\gamma\omega)$ with parameters $\epsilon_{\infty}^{}$ = 6.7, $\omega_l^{}$ = 1.827 $\times$ 10$^{14}$ rad$\cdot$s$^{-1}$, $\omega_t^{}$ = 1.495 $\times$ 10$^{14}$ rad$\cdot$s$^{-1}$, and $\gamma$ = 0.9 $\times$ 10$^{12}$ rad$\cdot$s$^{-1}$ \cite{Palik}.

\begin{figure} [htbp]
     \centering
     \subfigure [Insulator-phase VO$_2$] {\includegraphics[width=0.45\textwidth]{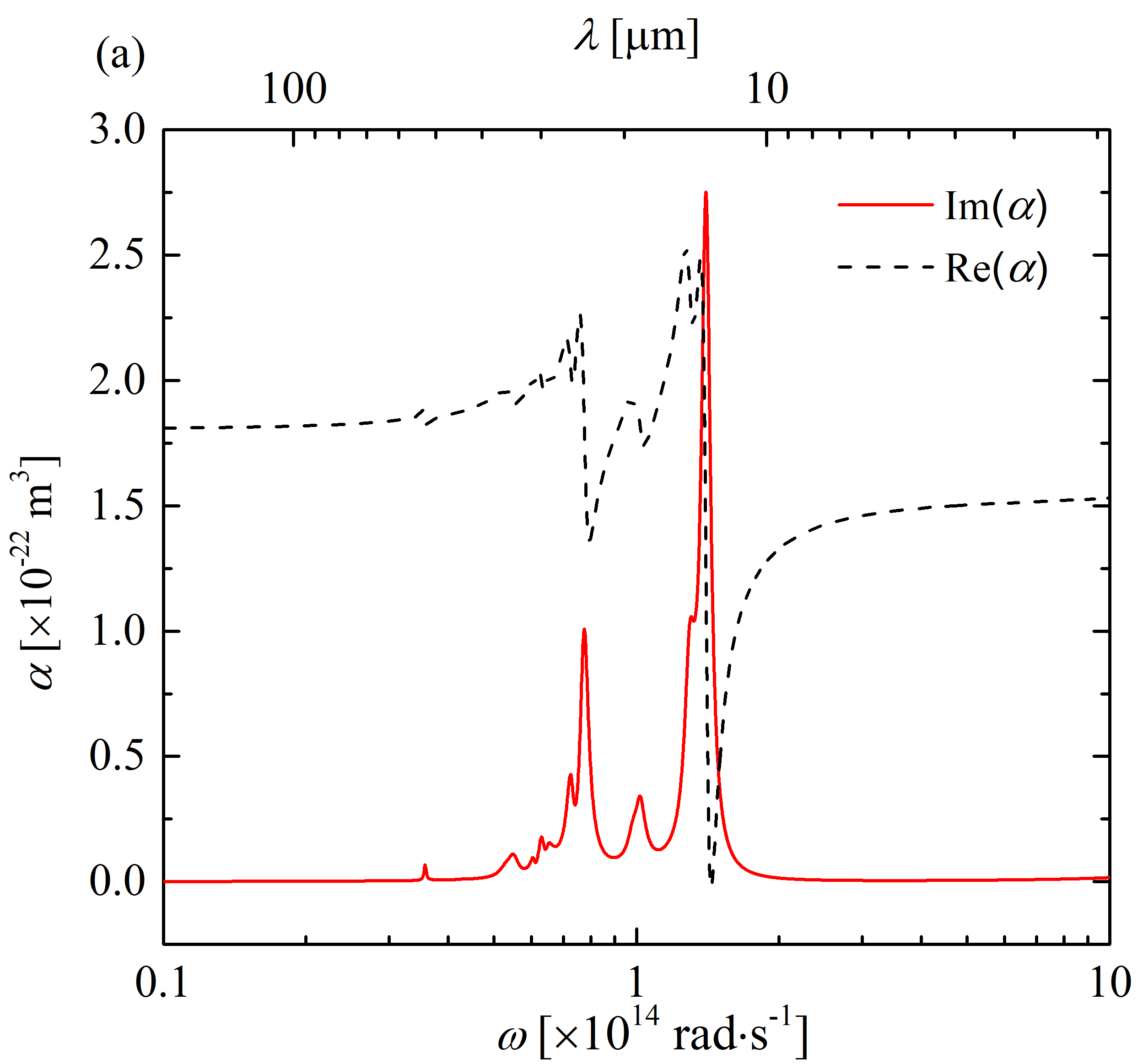}}
     \hspace{8pt}
     \subfigure [metallic-phase VO$_2$] {\includegraphics[width=0.48\textwidth]{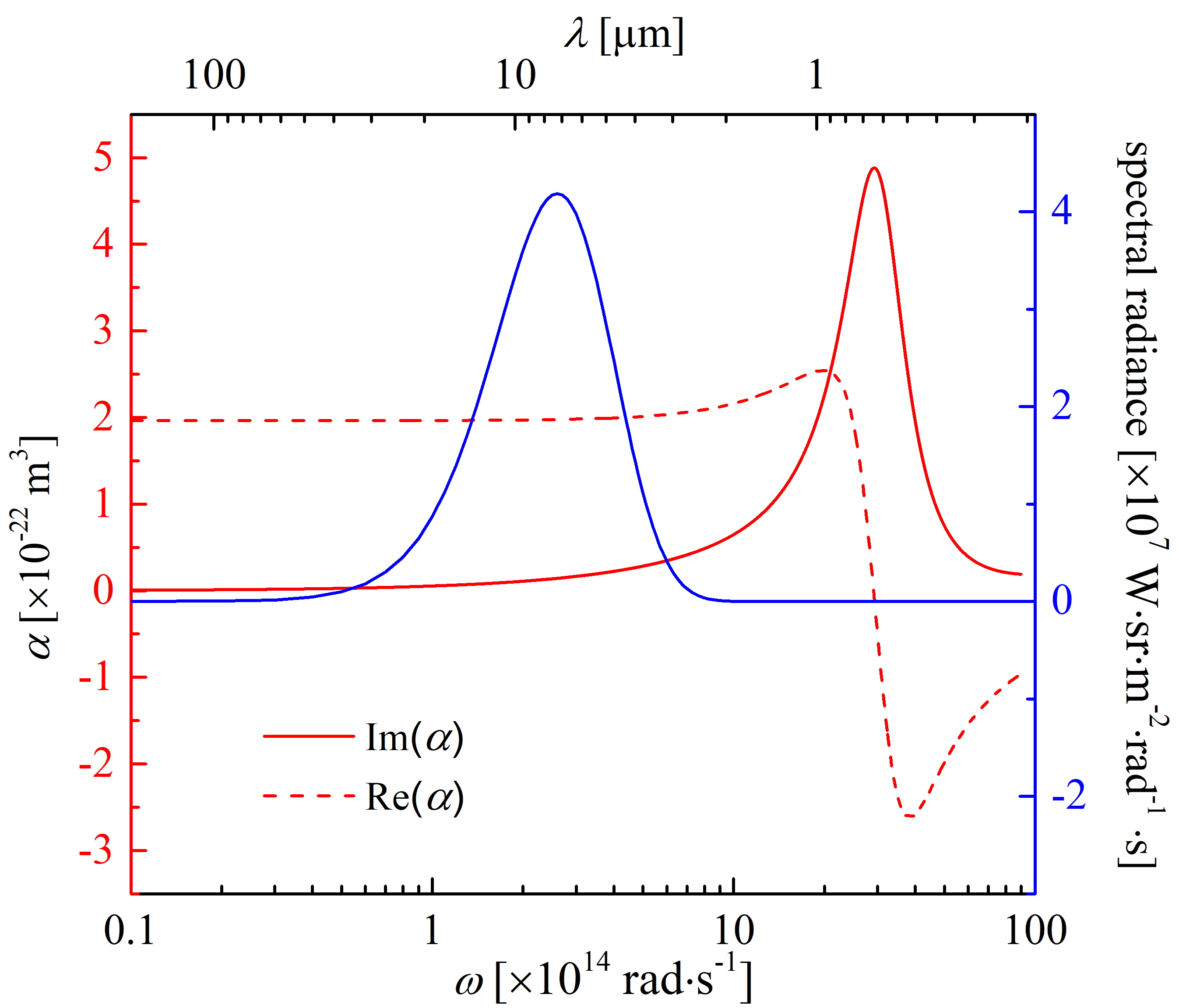}}
        \caption{Polarizability of VO$_2$ nanoparticle: (a) insulator phase and (b) metallic phase. Nanoparticle radius $a$ is 25 nm. $\epsilon_{\rm m}=1$. For insulator VO$_2$ particle, the ``$1/ 3-2/ 3$'' rule is applied to calculate the polarizability with the help of $\epsilon_\rVert^{}$ and $\epsilon_\bot^{}$ \cite{Quinten2010}. The spectral radiance of the blackbody at 400 K is also added for reference.}
        \label{polarizability}
\end{figure}

For an isotropic material embedded in the host medium with $\epsilon_{\rm m}$, the polarizability can be obtained from the first order Lorenz-Mie scattering coefficient \cite{Bohren1983,Mulholland1994}.
\begin{equation}
\alpha_E^{}=\frac{i6\pi}{k^3}a_1^{},
\label{Mie_E}
\end{equation}
where $a_1^{}$ is the first order Lorenz-Mie scattering coefficient defined as

\begin{equation}
a_1^{}=\frac{\epsilon/\epsilon_{\rm m} j_1^{}(y)[xj_1^{}(x)]\textquotesingle-j_1^{}(x)[yj_1^{}(y)]\textquotesingle}{\epsilon/\epsilon_{\rm m} j_1^{}(y)[xh_1^{(1)}(x)]\textquotesingle-h_1^{(1)}(x)[yj_1^{}(y)]\textquotesingle},
\label{Mie_scattering_coefficient}
\end{equation}
where $x=ka$, $y=\sqrt{\epsilon/\epsilon_{\rm m}}ka$, $a$ is the nanoparticle radius, $\epsilon$ is the relative permittivity, $j_1^{}(x)=\sin(x)/x^2-\cos(x)/x$ and $h_1^{(1)}(x)=\textrm{e}^{ix}(1/ix^2-1/x)$ are the first order Bessel function and spherical Hankel function. For 1D nanoparticle chains composed of many anistropic insulator-phase VO$_2$ nanoparticles, we assume that nanoparticles’ anisotropic axes are randomly oriented. For this reason, we decide to use the well-known $1/ 3-2/ 3$ description given in Ref.~\cite{Quinten2010} and consisting of two steps: first calculate polarizability for nanoparticle using $\epsilon_\bot^{}$ and $\epsilon_\rVert^{}$ separately, and then add up the results according to the $1/ 3-2/ 3$ rule \cite{Luo2020}:
\begin{equation}
\alpha_{E}^{}=\frac{1}{3}\alpha_{E}^{}(\epsilon_\rVert^{})+\frac{2}{3}\alpha_{E}^{}(\epsilon_\bot^{}).
\label{CM_ave}
\end{equation}
The polarizabilities for both insulator-phase and metallic-phase VO$_2$ nanoparticles are shown in Fig.~\ref{polarizability}. In order to compare the resonance frequency to the characteristic thermal frequency, the spectral radiance of the blackbody at 400 K is also added in Fig.~\ref{polarizability}(b) for reference. The characteristic thermal frequency mismatches with the polarizability resonance frequency of metallic VO$_2$ nanoparticle.

\begin{figure} [htbp]
\centerline {\includegraphics[width=0.55\textwidth]{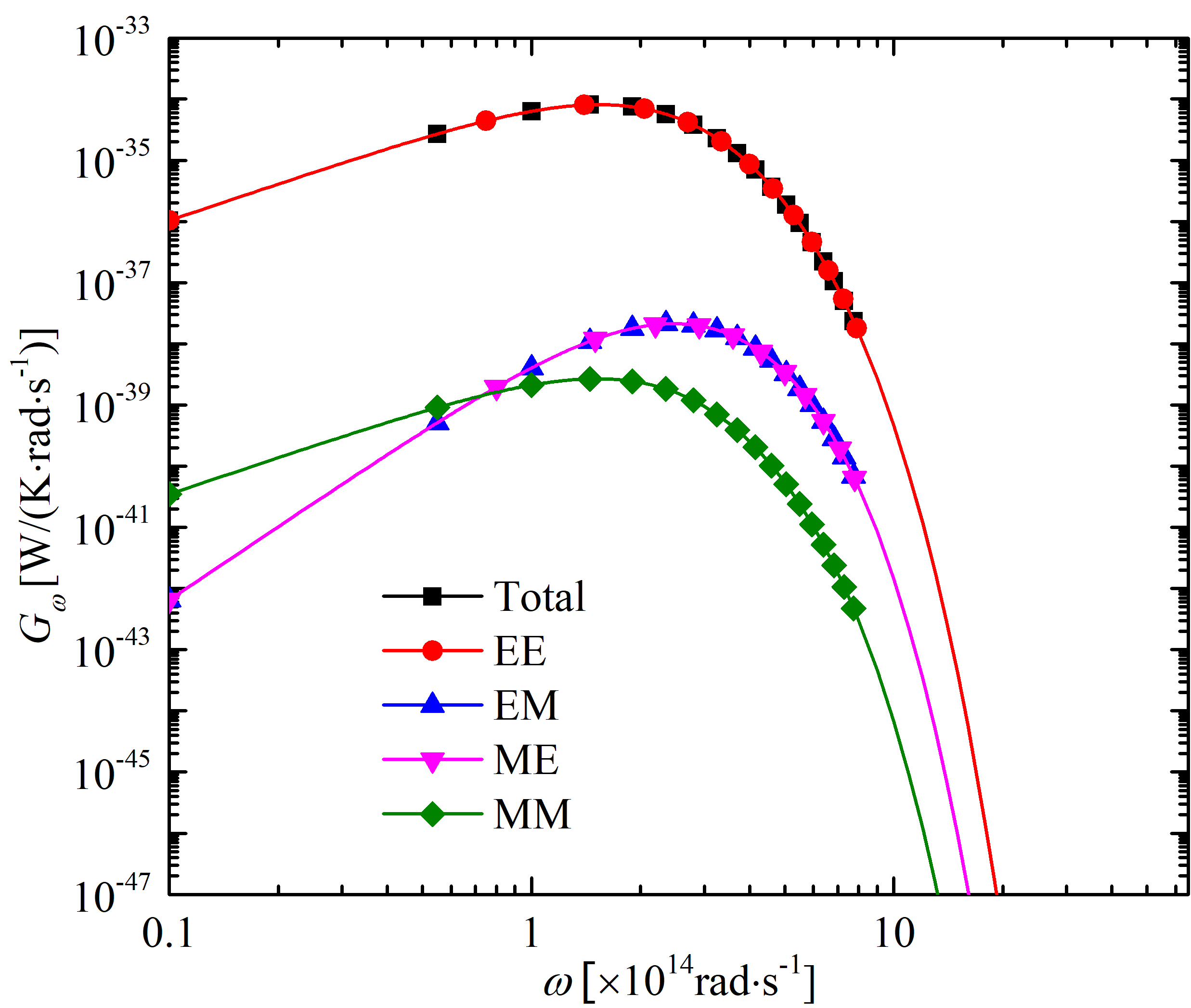}}
\caption{Spectral thermal conductance between two metallic-phase VO$_2$ nanoparticles, $G_\omega$, due to EE, EM, ME and MM contribution, respectively. Nanoparticle radius $a$ is 20 nm, temperature $T$ is 300 K. Separation between the two nanoparticle is 500 nm center to center.}
\label{four_contributions}
\end{figure}

For the metallic-phase VO$_2$, the spectral thermal conductance $G_\omega$ between two metallic-phase VO$_2$ nanoparticles with a separation 500 nm center to center is shown in Fig.~\ref{four_contributions}. Nanoparticle radius $a$ is 20 nm. Temperature $T$ is 300 K. All the four contributions, i.e., electric-electric polarized displacement current dissipation (EE), magnetic-electric polarized eddy-current Joule dissipation contribution (ME contribution), electric-magnetic polarized displacement current dissipation
(EM contribution) and magnetic-magnetic polarized eddy-current Joule dissipation (MM) are calculated by the coupled electric and magnetic dipole approach \cite{Luo2020,DongPrb2017}. Unlike the simple metal Ag, it is still the EE contribution that dominates the total thermal conductance, rather than the MM contribution, as shown in the Fig.~\ref{four_contributions}. Hence, when calculating radiative thermal conductance between phase-change VO$_2$ nanoparticles even in metallic phase, we can consider only the electric dipole and neglect the magnetic dipole safely.

\subsection{Effect of the phase change on ETC}

\begin{figure} [htbp]
\centerline {\includegraphics[width=0.55\textwidth]{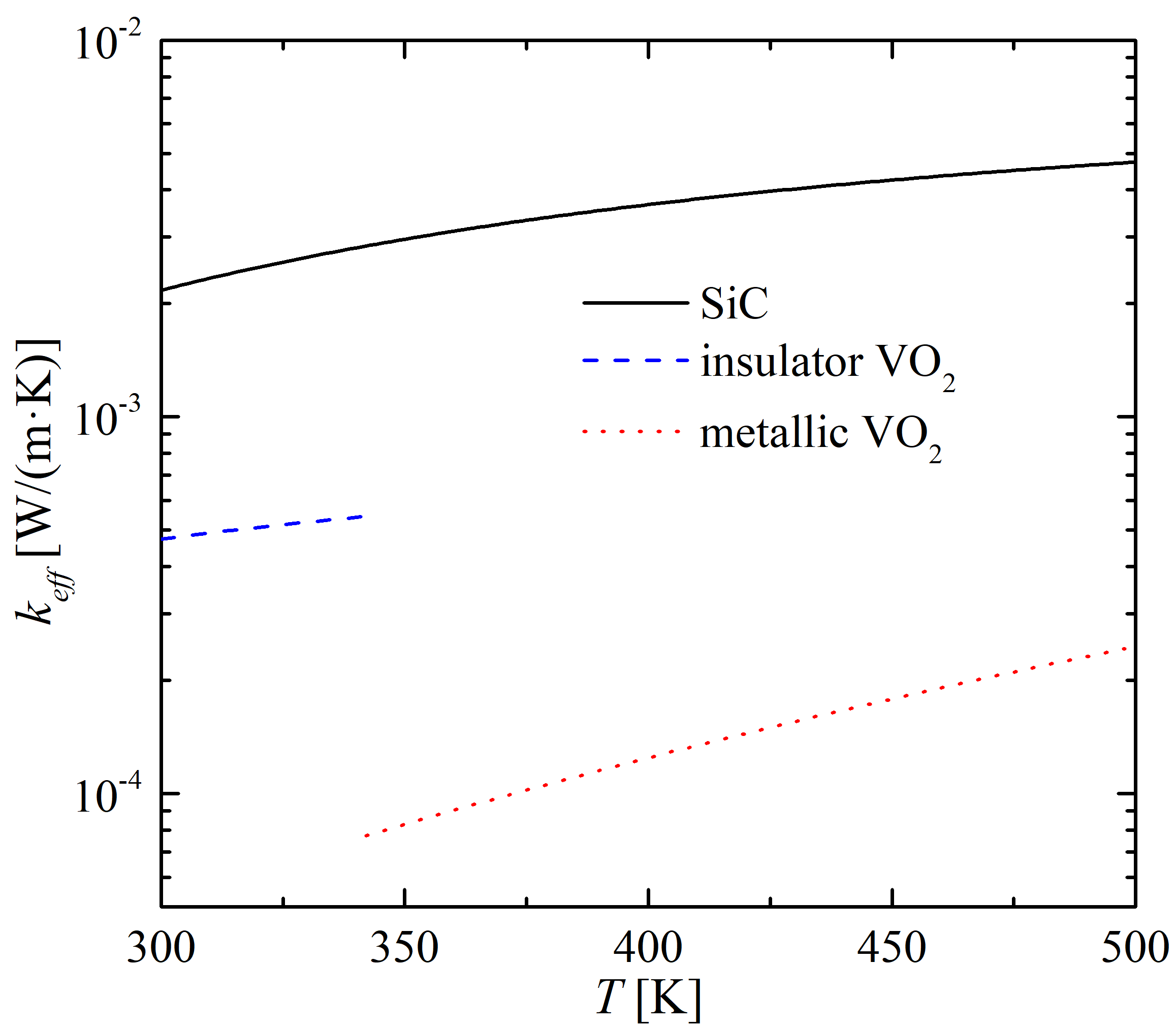}}
\caption{Effective thermal conductivity $k_{eff}$ of nanoparticle chains as a function of temperature. Phase-change VO$_2$ and polar SiC are considered. Nanoparticle radius $a$ = 25 nm. Lattice spacing $h$ =75 nm. $\epsilon_{\rm m}=5$.}
\label{phase_change}
\end{figure}

The ETC of the ordered nanoparticle chains (as shown in Fig.~\ref{Structure_diagram}) as a function of temperature $T$ is shown in Fig.~\ref{phase_change}. Here $a$ = 25 nm and $h$ = 75 nm. SiC and insulator-metal phase-change VO$_2$ nanoparticles chains are embedded in a host medium with $\epsilon_{\rm m}=5$. The temperature $T$ ranges from 300 K to 500 K, including the transition temperature of the VO$_2$.

For the non-phase-change SiC nanoparticle chain, the ETC increases monotonically with temperature. While for the insulator-metal phase-change VO$_2$ nanoparticle chain, an obvious transition of the ETC can be observed around the transition temperature of the VO$_2$. In the temperature range of interest, the ETC for the metallic-phase VO$_2$ nanoparticle chain at high temperature is even much lower than that of the insulator-phase VO$_2$ nanoparticle chain at low temperature, which is due to the insulator-metal phase change of VO$_2$. As shown in Fig.~\ref{polarizability}(b), an obvious mismatch between the resonance frequency of the metallic-phase VO$_2$ nanoparticle and the characteristic thermal frequency (Planck\textquotesingle s window), which accounts for the low ETC. However, for the insulator-phase VO$_2$ nanoparticles, the resonance frequency matches well with the characteristic thermal frequency as shown in Fig.~\ref{polarizability}(a), which accounts for the high ETC.

\begin{figure} [htbp]
\centerline {\includegraphics[width=0.55\textwidth]{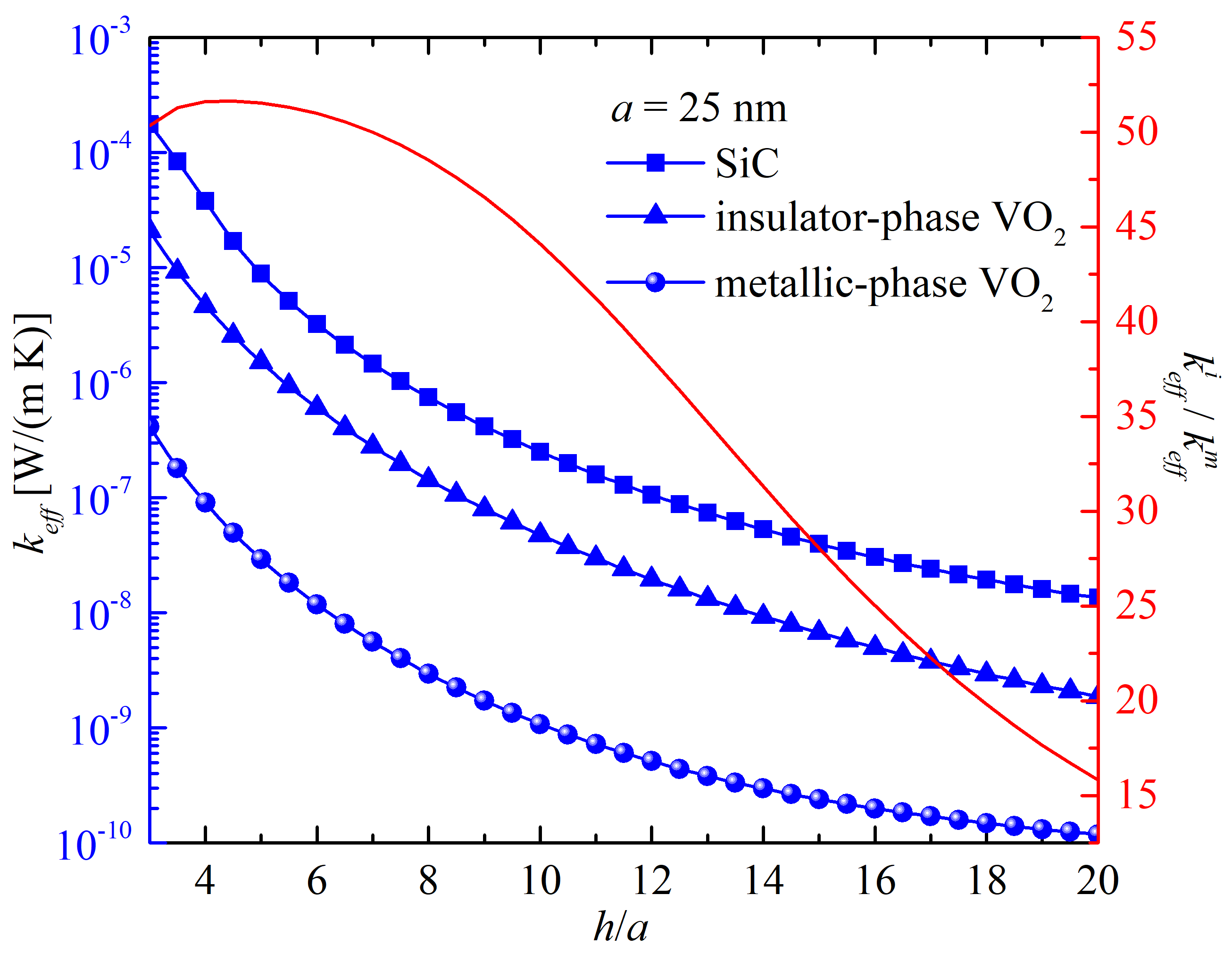}}
\caption{Dependence of the ETC on $h/a$. SiC (300K), insulator-phase VO$_2$ (300K) and metallic-phase VO$_2$ (400K) nanoparticles chains in vacuum are considered. The dependence of the ratio of $k_{eff}^{i}$ (the ETC for insulator-phase VO$_2$) to $k_{eff}^{m}$ (the ETC for metallic-phase VO$_2$) on $h/a$ is also added. Nanoparticle radius $a=25$ nm.}
\label{ETC_MBI}
\end{figure}

To give a quantitative description on the phase change effect, the dependence of the ETC on $h/a$ is shown in Fig.~\ref{ETC_MBI}. SiC (300K), insulator-phase VO$_2$ (300K) and metall-phase VO$_2$ (400K) nanoparticles chains in vacuum are considered. The dependence of the ratio of $k_{eff}^{i}$ (the ETC for insulator-phase VO$_2$) to $k_{eff}^{m}$ (the ETC for metallic-phase VO$_2$) on $h/a$ is also shown in Fig.~\ref{ETC_MBI}. In general, the ratio $k_{eff}^{i}/k_{eff}^{m}$ is much larger than unity. The ETC for the insulator-phase VO$_2$ nanoparticle chains is much larger than that of the metallic-phase VO$_2$ nanoparticle chains. The ratio $k_{eff}^{i}/k_{eff}^{m}$ increases to its maximum (around 50) and then decreases with increasing $h/a$. The phase change effect is significant when the chain is compact and decreases when the chain goes dilute. In addition, the ETC decreases with increasing $h/a$. The inter-particle coupling decreases when the lattice spacing of the nanoparticle chain $h$ increases. The decreasing coupling accounts for the decreasing ETC when increasing $h/a$.

\subsection{Effect of the many-body interaction on ETC}
To evaluate the effects of the many-body interaction on the ETC, we define the following parameter \cite{Luo2019}:

\begin{equation}
\varphi=\frac{k_{eff}^{}}{k_{eff}^{0}},
\label{D_MBI_ratio}
\end{equation}
where $k_{eff}$ is the ETC evaluated with the help of the radiative thermal conductance in Eq.(\ref{conductance}) and the Green\textquotesingle s function including the MBI in Eq.(\ref{Green_function_MBI}), $k_{eff}^{0}$ is the ETC without the MBI evaluated with the help of the Eq.(\ref{conductance}) and the free space Green\textquotesingle s function. Generally speaking, the MBI inhibits the ETC when $\varphi<1$, enhances it when $\varphi>1$ and can be neglected safely when $\varphi \approx 1$.

\begin{figure} [htbp]
\centerline {\includegraphics[width=0.6\textwidth]{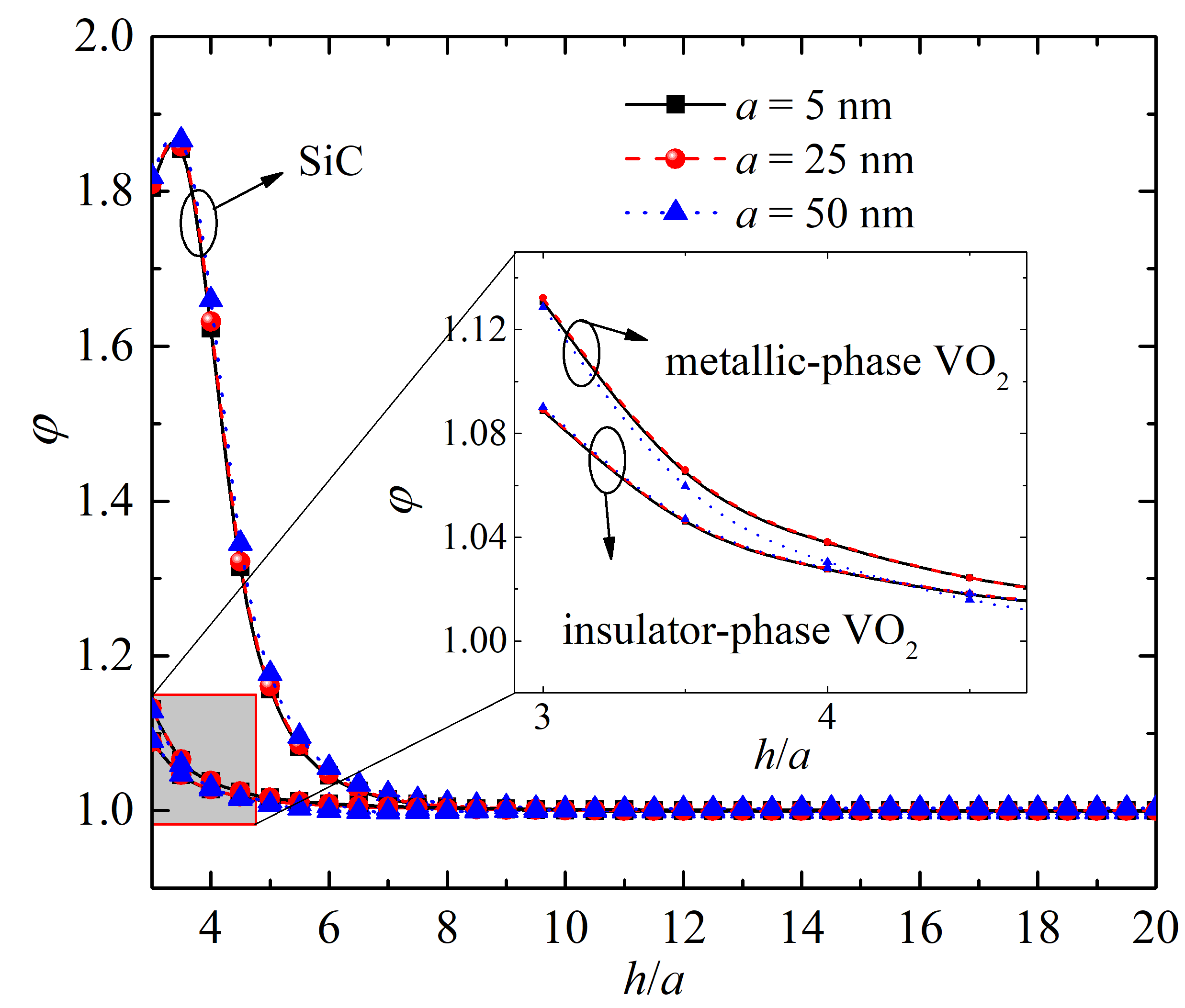}}
\caption{The dependence of the ratio $\varphi$ of the effective thermal conductivity $k_{eff}$ of nanoparticle chains with the MBI to that without MBI on $h/a$. Insulator-phase VO$_2$, metallic-phase VO$_2$ and polar SiC are considered. Nanoparticles of three different sizes have been considered $a$ = 5 nm, 25 nm and 50 nm. $\epsilon_{\rm m}=1$.}
\label{MBI_ETC}
\end{figure}

We quantitatively evaluate the MBI effect on the ETC for polar SiC, metallic-phase VO$_2$ and insulator-phase VO$_2$ nanoparticle chains. The dependence of the parameter $\varphi$ defined by Eq.(\ref{D_MBI_ratio}) on the geometrical dimensionless parameter $h/a$ is shown in Fig.~\ref{MBI_ETC}. Nanoparticles of three different sizes have been considered $a$ = 5 nm, 25 nm and 50 nm. As shown in Fig.~\ref{MBI_ETC}, $\varphi$ is never less than unity, which indicates that the MBI does not inhibit the ETC for the chains composed of the considered materials. When $h/a>8$, $\varphi$ starts to approach unity. The MBI decreases with the increasing lattice spacing $h$. When $h/a<8$, $\varphi>1$. The MBI is favorable to the ETC. Small lattice spacing (i.e., $h/a<8$) results in strong inter-particles couplings, which resulting in a significant MBI.

The maximal $\varphi$ for SiC nanoparticle chains is around 2. The MBI can double the ETC for SiC chains. For SiC nanoparticle chains, an extremum value for $\varphi$ was observed at $h/a\approx 3.5$, of which the insight is still unclear and remains to be explored in the future. It\textquotesingle s worth mentioning that a similar extremum value for ratio of radiative heat flux between two nanoparticles with insertion of a third nanoparticle to that of the two isolated nanoparticles when increasing the distance between the two nanoparticles has already been reported \cite{Ben2011}. It\textquotesingle s also observed in Fig.~\ref{MBI_ETC} that $\varphi$ does not change with varying the nanoparticle size. That is to say that the MBI is independent of the nanoparticle size. $\varphi$ for the metallic-phase VO$_2$ and insulator-phase VO$_2$ nanoparticle chains is similar to each other, though $\varphi$ for the metallic-phase VO$_2$ chains is a little bit larger than that of the insulator-phase VO$_2$ chains.

\subsection{Effect of the host medium relative permittivity on ETC}

\begin{figure} [htbp]
\centerline {\includegraphics[width=0.6\textwidth]{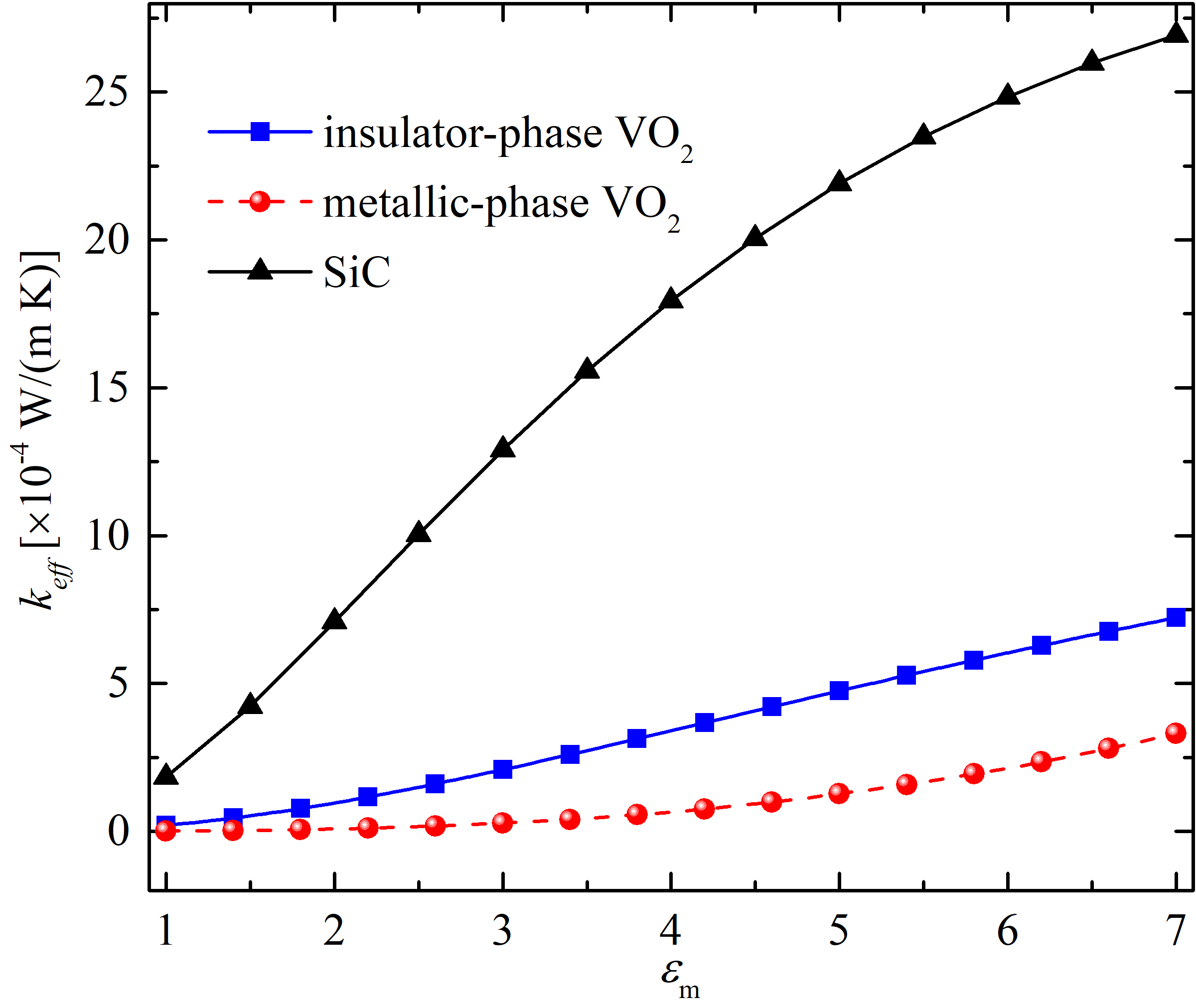}}
\caption{The dependence of ETC on the host medium permittivity $\epsilon_{\rm m}$ for insulator-metal phase-change VO$_2$ and SiC. Insulator-phase VO$_2$ (300 K), metallic-phase VO$_2$ (400 K) and SiC (300 K) are considered. Nanoparticle radius $a$ = 25 nm. $h=75$ nm.}
\label{epsm_effect}
\end{figure}

The dependence of total ETC on the host medium permittivity $\epsilon_{\rm m}$ for insulator-metal phase-change VO$_2$ and SiC is show in Fig.~\ref{epsm_effect}. Nanoparticles with radius $a$ = 25 nm and $h=75$ nm are used. ETC increases with the host medium relative permittivity $\epsilon_{\rm m}$ for both insulator-metal phase-change VO$_2$ and SiC. The host medium relative permittivity significantly affects the inter-particle coupling, which finally significantly affects the ETC. High relative permittivity is favorable to enhancing the ETC and radiative heat transfer in the nanoparticle chain, which is consistent with the reported results for closely spaced metallic nanoparticle chains \cite{Ben2008}. In addition, at low relative permittivity $\epsilon_{\rm m}$, the difference between the ETC for the VO$_2$ nanoparticle chains at different phases (i.e., metallic phase and insulator phase) is much smaller than that at high relative permittivity, of which the insight will be analyzed in the following from the thermal conductivity spectrum standpoint.

\begin{figure} [!htbp]
     \centering
     \subfigure [Insulator-phase VO$_2$] {\includegraphics[width=0.485\textwidth]{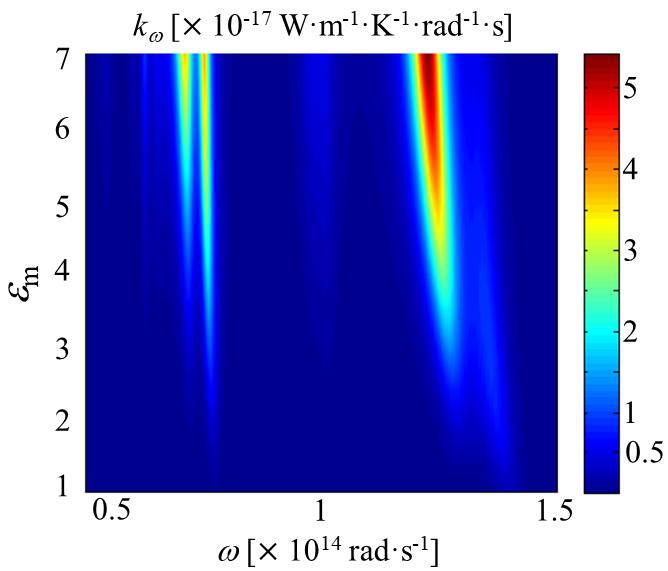}}
     \hspace{8pt}
     \subfigure [metallic-phase VO$_2$] {\includegraphics[width=0.48\textwidth]{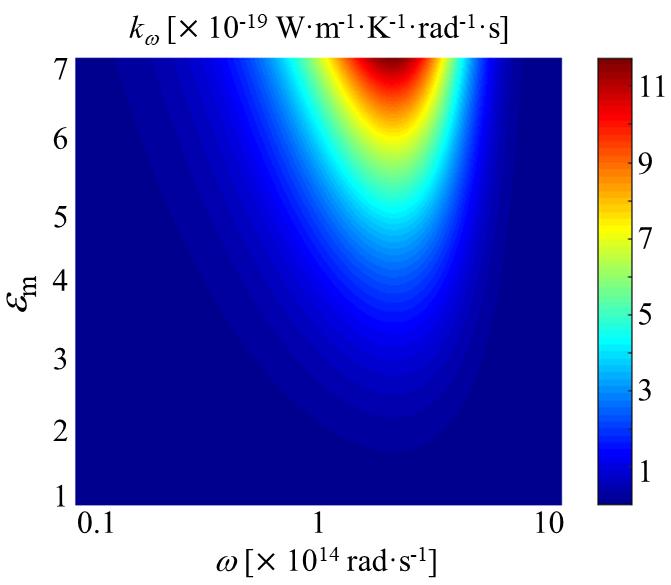}}
        \caption{Dependence of the spectral effective thermal conductivity $k_{\omega}$ on the $\omega$ and relative permittivity of the host medium $\epsilon_{m}$: (a) insulator phase (300 K) and (b) metallic phase (400 K). Nanoparticle radius $a$ is 25 nm. $h=75$ nm.}
        \label{k_epsilon}
\end{figure}

The dependence of the spectral effective thermal conductivity $k_{\omega}$ on the angular frequency $\omega$ and the relative permittivity of the host medium $\epsilon_{\rm m}$ is shown in Fig.~\ref{k_epsilon}: (a) insulator-phase VO$_2$ (300 K) and (b) metallic-phase VO$_2$ (400 K) ($a=25$ nm and $h=75$ nm). For a fixed angular frequency, the value of the spectral effective thermal conductivity increases significantly with increasing $\epsilon_{\rm m}$, which is consistent with the dependence of total ETC on $\epsilon_{\rm m}$, as shown in Fig.~\ref{epsm_effect}. Increasing the relative permittivity is in favor of enhancing the radiative effective thermal conductivity. For metallic-phase VO$_2$ nanoparticle chains, the frequency peak of the spectral effective thermal conductivity corresponds to the characteristic thermal frequency, as shown in Fig.~\ref{k_epsilon}(b). However, for insulator-phase VO$_2$ nanoparticle chains, besides the peak of the spectral effective thermal conductivity corresponding to the characteristic thermal frequency, there are several secondary peaks, as shown in Fig.~\ref{k_epsilon}(a). It\textquotesingle s worthwhile to mention that the peak of the spectral effective thermal conductivity shows a red-shift behavior with the increase of relative permittivity $\epsilon_{\rm m}$ for both insulator-phase and metallic-phase VO$_2$, as can be seen from Fig.~\ref{k_epsilon}.

\begin{figure} [!htbp]
     \centering
     \subfigure [Insulator-phase VO$_2$] {\includegraphics[width=0.475\textwidth]{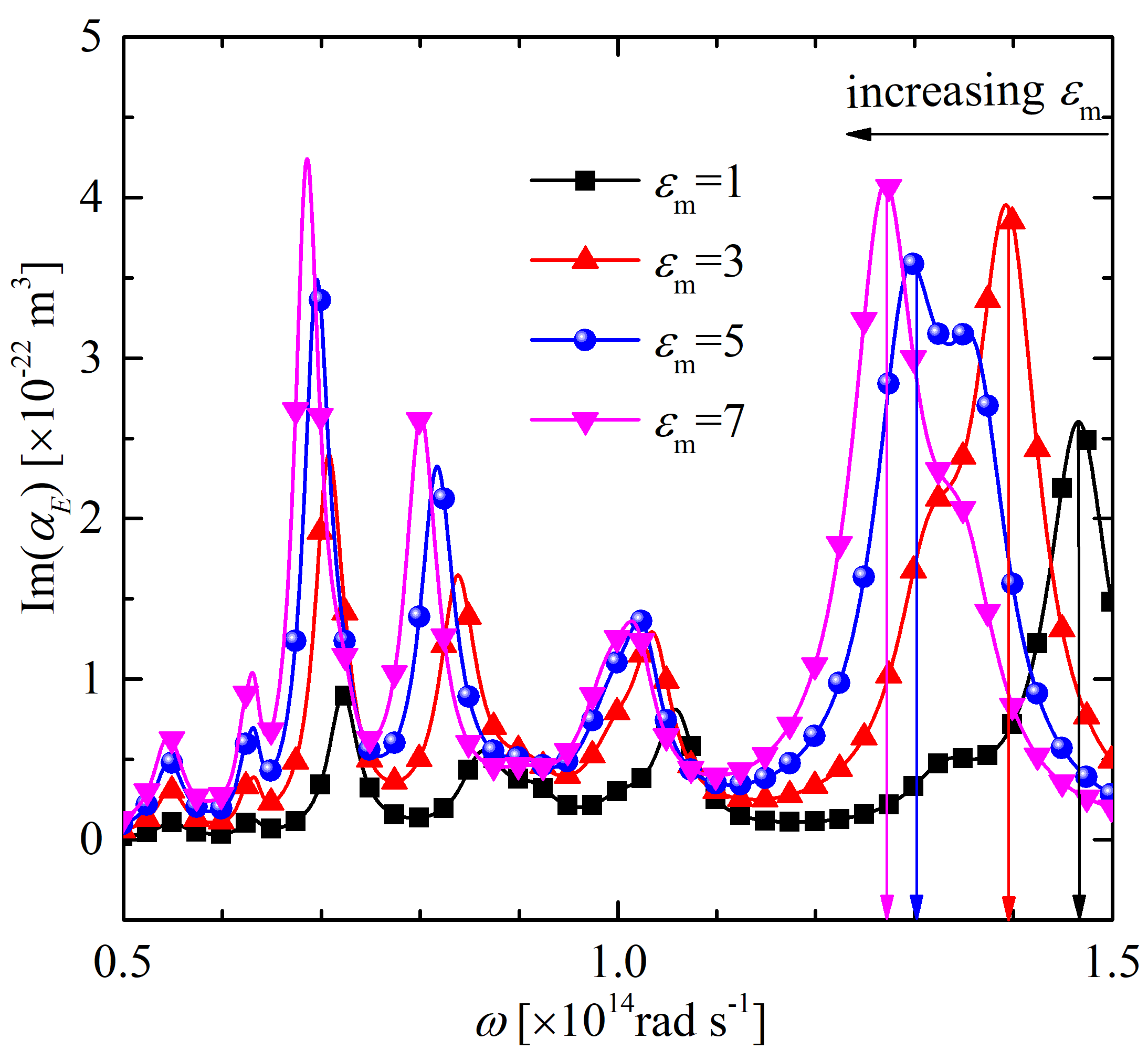}}
     \hspace{8pt}
     \subfigure [metallic-phase VO$_2$] {\includegraphics[width=0.48\textwidth]{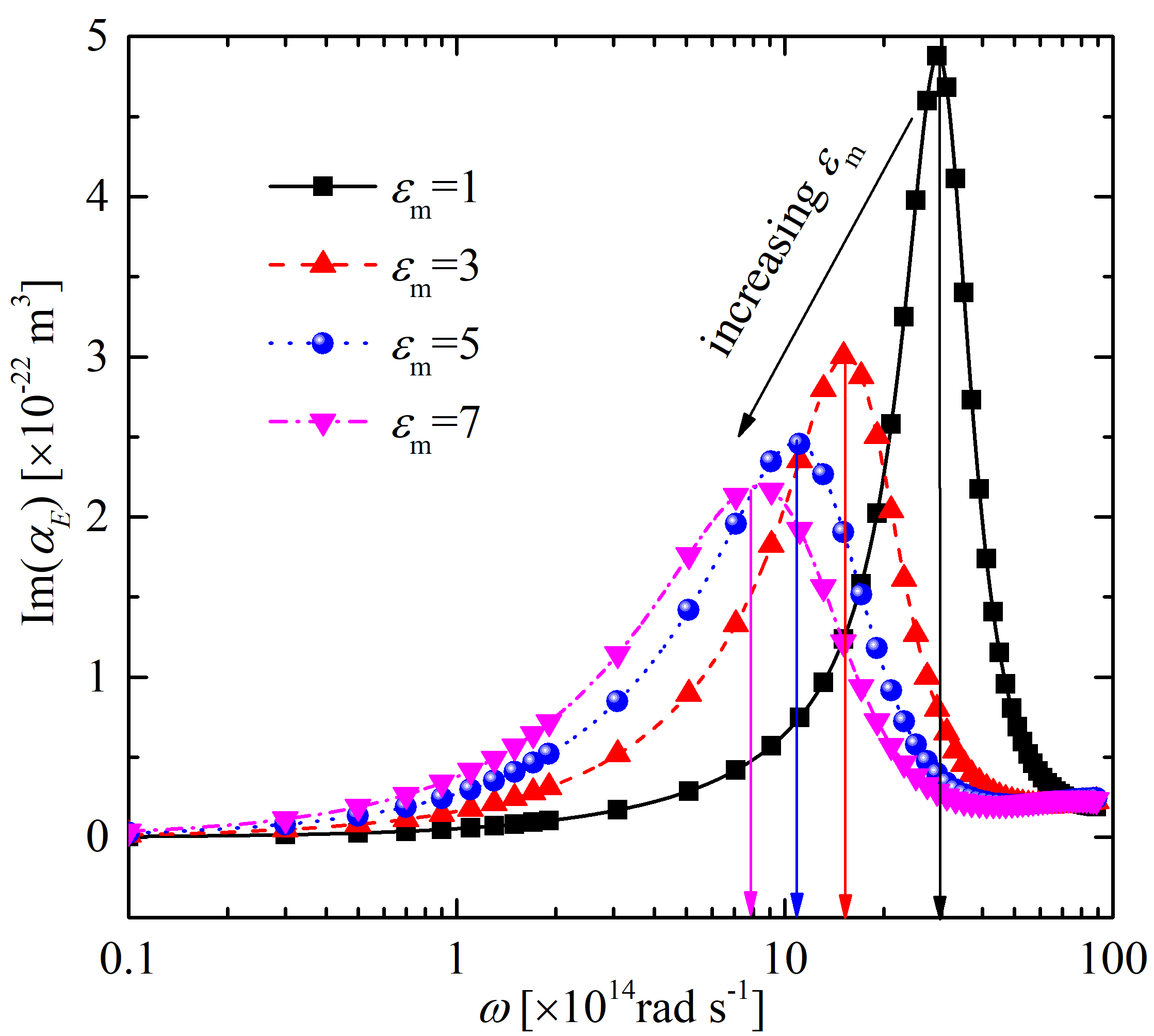}}
        \caption{Polarizability of VO$_2$ nanoparticle at several different relative permittivities $\epsilon_{\rm m}$ : (a) insulator phase and (b) metallic phase. Nanoparticle radius $a$ is 25 nm. Relative permittivity $\epsilon_{\rm m}$= 1, 3, 5 and 7, respectively. The increasing directions of the $\epsilon_{\rm m}$, as well as the angular frequency corresponding to the main peaks, are added for reference.}
        \label{Polarizability_VO2_epsm}
\end{figure}

From the formulas for the ETC (i.e., Eq.~(\ref{ETC_rearranged}) combined with the Eq.~(\ref{conductance})), the polarizability for the nanoparticle plays a significant role in determining the ETC for the nanoparticle chains. To understand the insight of the red-shift behavior of the peaks with increasing the relative permittivity $\epsilon_{\rm m}$ and different spectral behaviors of the effective thermal conductivity for the insulator-phase VO$_2$ and metallic-phase VO$_2$ nanoparticle chains, the polarizability of the VO$_2$ nanoparticles embedded in the host medium with several different relative permittivities $\epsilon_{\rm m}$ is given in Fig.~\ref{Polarizability_VO2_epsm} (a) for insulator-phase  VO$_2$ nanoparticles and (b) for metallic-phase VO$_2$ nanoparticles with the following parameters. Nanoparticle radius $a$ = 25 nm. Relative permittivity $\epsilon_{\rm m}$= 1, 3, 5 and 7, respectively. The increasing directions of the $\epsilon_{\rm m}$, as well as the angular frequency corresponding to the main peaks, are also added for reference.

As shown in Fig.~\ref{Polarizability_VO2_epsm}, the peak of the polarizability shows an obvious red-shift behavior for both the insulator-phase VO$_2$ and metallic-phase VO$_2$ nanoparticles, which accounts for the red-shift behaviors of the peaks for the spectral effective thermal conductivity as shown in Fig.~\ref{k_epsilon}. This red-shift behavior of the peaks for the polarizability with increasing $\epsilon_{\rm m}$ results in the increasing match between the peak frequency of the polarizability and the characteristic thermal frequency (Planck\textquotesingle s window, as shown in Fig.~\ref{polarizability}(b)), which accounts for the increasing total ETC with increasing $\epsilon_{\rm m}$ as shown in Fig.~\ref{epsm_effect}.

The main reason for the observed increasing ETC of the VO$_2$ chain in the metallic phase when changing $\epsilon_{\rm m}$ is that the resonance is red-shifted so that the heat flux is enhanced in the infra-red regime by the low frequency side of the resonance. This red-shift for the resonances is also known for simple metallic (Drude model $\epsilon=\epsilon_{\infty}-\omega_{p}^2/{\omega}^2$) nanoparticles where the localized resonances are given by $\omega_{p}/\sqrt{\epsilon_{\infty}+ 2 \epsilon_{\rm m}}$. For the metallic-phase VO$_2$ nanoparticle chain, we also can observe a competition between the following two processes: (1) the decreasing peak value of the metallic-phase VO$_2$ nanoparticle polarizability with increasing $\epsilon_{\rm m}$ and (2) the increasing match degree between the peak frequency of the polarizability and the characteristic thermal frequency (Planck\textquotesingle s window) with increasing $\epsilon_{\rm m}$. As shown in Fig.~\ref{epsm_effect}, the total ETC increases with increasing the $\epsilon_{\rm m}$, therefore, the match degree between the peak frequency of the polarizability and the characteristic thermal frequency (Planck\textquotesingle s window) is the influencing factor prior to the exact value of the polarizability peak.

\begin{figure} [htbp]
\centerline {\includegraphics[width=0.6\textwidth]{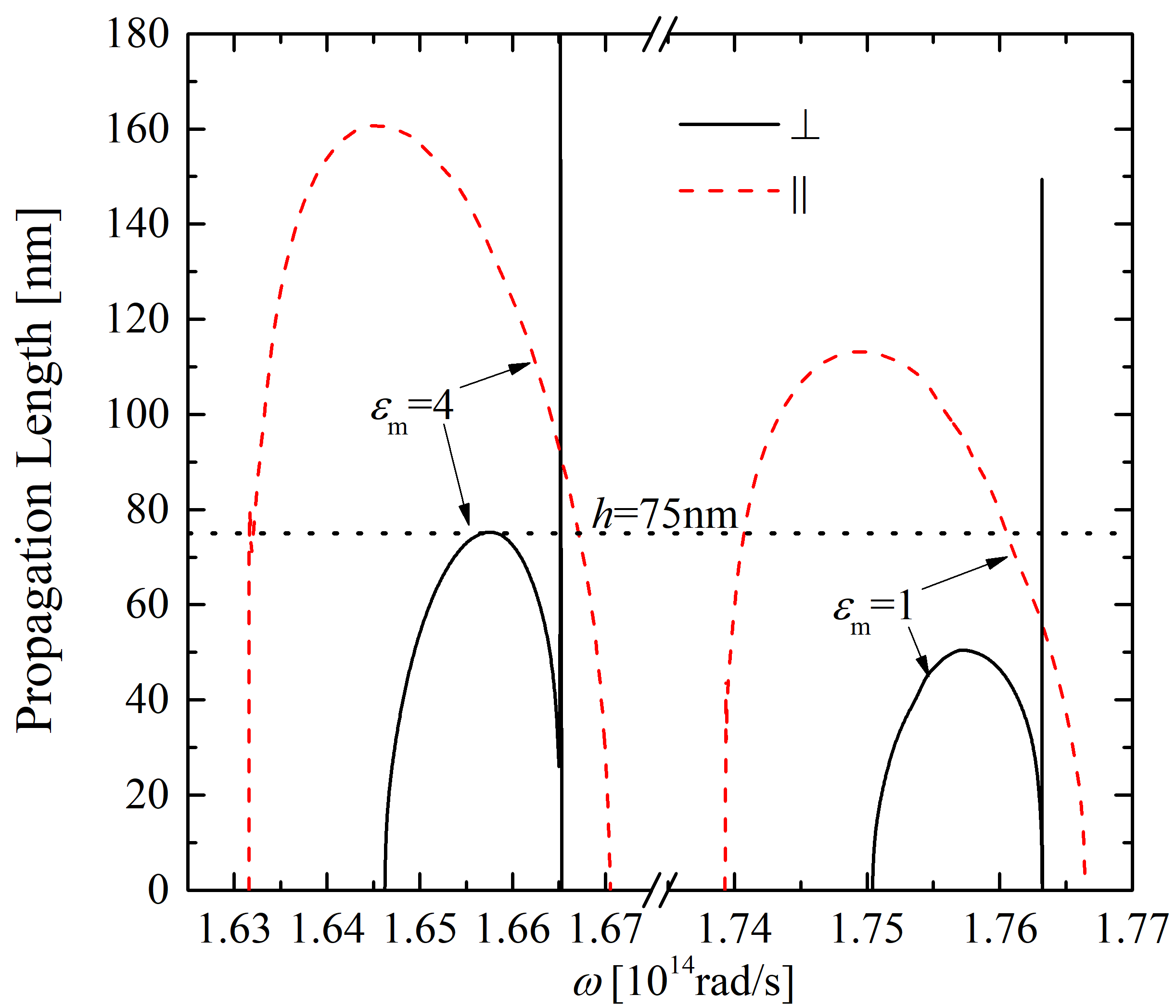}}
\caption{The propagation length of the SiC nanoparticle chain. Two host medium permittivities are considered, $\epsilon_{\rm m} =1$ and 4, respectively. Nanoparticle radius $a$ = 25 nm. Lattice spacing $h=75$ nm. The symbols $\rVert$ and $\bot$ are corresponding to the parallel and perpendicular polarizations, respectively.}
\label{propagation_length}
\end{figure}

As shown in Fig.~\ref{epsm_effect}, for the materials supporting coupled resonant modes in the infra-red regime (i.e., SiC and metallic-phase VO$_2$), increasing the host medium permittivity will significantly enhance the ETC as compared to the materials supporting no coupled resonant modes in the infra-red regime. The main reason is that the resonances modes in the infra-red regime are significantly enhanced when increasing the host medium permittivity, which can be characterized by the increased propagation length defined as $\frac{{\rm d}{\rm Re}(\omega)}{{\rm d}k}\frac{1}{2{\rm Im}(\omega)}$ \cite{Kathmann2018}. We take the simple dielectric SiC as an example to show the effect of the permittivity on the propagation length of coupled modes in infra-red regime. The dependence of the propagation length on the angular frequency for the SiC nanoparticle chain is shown in Fig.~\ref{propagation_length}. Nanoparticle radius $a=$ 25 nm. Lattice spacing $h=$ 75 nm. Two different host medium permittivities are considered, $\epsilon_{\rm m}=1$ and 4, respectively. When increasing the permittivity of the host medium, the propagation length increases significantly. Hence more energy can be transported and the ETC is significantly enhanced by the increased coupled resonant modes in infra-red regime. In addition, the red-shift of the corresponding resonant frequency of the coupled modes also can be observed in the Fig.~\ref{propagation_length}, when increasing the host medium permittivity $\epsilon_{\rm m}$.

\section{Conclusion}
Near-field radiative heat transfer for 1D nanoparticle chains embedded in a non-absorbing host medium is investigated from the point view of the continuum by means of the MF method combining the many-body radiative heat transfer theory and the Fourier law together. Effects of the phase change of materials, complex many-body interaction and host medium relative permittivity on the effective thermal conductivity ETC are analyzed.

The value of the ETC for VO$_2$ nanoparticle chains below the transition temperature can reach around 50 times that above the transition temperature due to the phase change effect. The strong coupling in the insulator-phase VO$_2$ nanoparticle chain accounts for its high ETC as compared to the low ETC for the metallic-phase VO$_2$ nanoparticle chain, where there is a mismatch between the characteristic thermal frequency and the polarizability resonance frequency.

Strong MBI is in favor of the ETC. For dense chains (the ratio of the lattice spacing to nanoparticle radius $h/a<8$), the MBI enhances the ETC, which is due to the strong coupling in the dense chains. When the chains go more and more dilute ($h/a>8$), the MBI can be neglected safely, which is due to negligible coupling. It\textquotesingle s worthwhile to mention that for the SiC nanoparticle chain the MBI can even double the ETC of the chain as compared to that without considering MBI. It is still remaining unknown whether there are some other materials in nature supporting an even larger MBI effect on ETC than SiC.

The host medium relative permittivity significantly affects the inter-particle coupling, which accounts for the permitivity-dependent ETC for the VO$_2$ nanoparticle chains. For the materials supporting no resonant modes in the infra-red regime (e.g., simple metal or metallic-phase VO$_2$), the red-shift behavior of the peaks for the polarizability with increasing $\epsilon_{\rm m}$ results in the increasing degree of match between the peak frequency of the polarizability and the characteristic thermal frequency (Planck\textquotesingle s window), which accounts for increasing the total ETC with increasing $\epsilon_{\rm m}$. For the materials supporting resonant modes in the infra-red regime (e.g., simple dielectric SiC and insulator-phase VO$_2$), more energy can be transported and then ETC can be enhanced due to the increased coupled resonance when increasing $\epsilon_{\rm m}$.

It is noted that the topological phase transition has a significant effect on the radiative heat flux through the 1D chain of nanoparticles \cite{Ott2020}. However, the effect of such topological phase transition on the thermal transport characteristics (e.g., ETC) still remains unclear and needs to be investigated in the future.

\section*{Acknowledgements} 
The support of this work by the National Natural Science Foundation of China (No. 51976045) is gratefully acknowledged. M.A. acknowledges support from the Institute Universitaire de France, Paris, France (UE). M.G.L. also thanks for the support from China Scholarship Council (No.201906120208).

\addcontentsline{toc}{section}{Acknowledgements}



\providecommand{\noopsort}[1]{}\providecommand{\singleletter}[1]{#1}%

\end{document}